# Transport Properties of Granular High-$T_C$ Superconductors

M. S. da Luz, C. A. M. dos Santos*, M. J. R. Sandim, and A. J. S. Machado

*Departamento de Engenharia de Materiais, FAENQUIL, Caixa Postal 116, 12600-970, Lorena-SP, Brazil*

R. F. Jardim

*Instituto de Física, Universidade de São Paulo, Caixa Postal 66318, 05315-970, São Paulo, SP, Brazil*

## ABSTRACT

We report on the application of the Resistively Shunted Junction (RSJ) model to granular high-$T_C$ superconductors. Some derived predictions of the RSJ model are applied to a set of superconducting granular samples which can be considered as a network of Josephson junctions. The investigated samples belong to both hole-doped $Y_{1-x}Pr_xBa_2Cu_3O_{7-\delta}$ (x = 0.0, 0.35, and 0.45) and the electron-doped $Sm_{2-x}Ce_xCuO_{4-\delta}$ (x = 0.18) systems which display the so-called double resistive superconducting transition. We have performed several transport measurements in these compounds including temperature and magnetic field dependence of the electrical resistance, R(T,H), and I-V characteristics. Several aspects of the I-V characteristics were quantitatively well described by the RSJ model. The combined results strongly suggest that dissipation in granular superconducting samples is a natural consequence of the normal current flowing in parallel with the supercurrent current.

*Corresponding author. Fax: +55-12-553-3006

*E-mail address*: cams@demar.faenquil.br

## I. INTRODUCTION

In the last years, much attention has been dedicated to understanding aspects related to superconductivity in granular and disordered systems [1-12]. One remarkable feature of granular superconductors is the occurrence of a double resistive superconducting transition [1-7]. Such a double transition is noticeable by two well defined drops in the electrical resistance of a given superconducting material at two distinct temperatures labeled $T_{ci}$ and $T_{cj}$. The first superconducting transition, which occurs at an upper temperature $T_{ci}$ is attributed to the development of superconductivity in isolated superconducting islands within the material. The second transition, which eventually drives the system to the zero resistance state, occurs at $T_{cj} < T_{ci}$. Such a transition is related to the long-range phase coherence of the order parameter in which Josephson coupling between superconducting islands is achieved [1,3,5,6]

Within the scenario of disordered superconducting systems it has been suggested that the I-V characteristics shows a power-law behavior,

$$V \propto (I - I_c)^a \qquad (1)$$

above a critical current Ic [10,11]. When *a* is equal to the unity, Eq. (1) is related to two possible mechanisms: *i*) flux flow dissipation [13,14], where $I_C$ is the maximum current in which vortices stay pinned and the Lorentz is not enough to overcome the pinning force, or *ii*) dissipation due to a normal current flowing in parallel with a supercurrent, as predicted by resistively shunt model RSJ model [8,9].

According to several authors [8,9,15-18], the dissipation in superconducting junctions is well described by the RSJ model in which a normal current can flow in parallel with a supercurrent, as schematized in the circuit displayed Figure 1. In such a circuit, when the current exceeds the critical current of the junction $I_C$ a normal current flows in parallel with the supercurrent and a dissipation process starts. Within this context, several works have been reported concerning the RSJ model applied to single grain-boundary Josephson junctions [16-18].

The RSJ model states that the total applied current I which crosses a junction is given by:

$$I = I_S + I_N, \quad (2)$$

where $I_S$ is the supercurrent and $I_N$ is the normal current related to the dissipation in the junction. The normal current is predicted to be described by

$$I_N = V / R_N, \quad (3)$$

where $R_N$ is a normal state shunt resistance which is assumed to be magnetic field independent, as displayed in Fig. 1. Combining Eqs. (2) and (3) yields

$$V = R_N (I - I_S). \quad (4)$$

There are few points to be addressed in Eq. 4. First, when the total applied current is increased, the maximum supercurrent through the junction is reached at $I_S = I_C$, where $I_C$ is the critical current of the junction. For $I_S = I_C$ and using Eq. (4) one is able to obtain

$$V = R_N (I - I_C), \quad (5)$$

which is the same as Eq. 1 when $a = 1$ [10,11].

Considering I-V characteristics measured at different applied magnetic fields (H) and fixed temperatures, it is of interest to analyze the consequences of Eq. (5). When H is lower than the upper intergranular critical field ($H_{c2J}$) [19-21], $I_C \neq$ zero and this indicates that a linear regime ($dV/dI = R_N$ = constant) at high excitation currents ($I > I_C$) may be observed in I-V characteristics. Furthermore, for applied magnetic fields $H > H_{c2J}$, $I_C$ is reduced to zero and a linear (ohmic) regime given by

$$V = R_N I, \qquad (6)$$

must be reached. In addition to this one would expect a saturation of the magnetoresistance of the system starting when $R = R_N$.

We mention that these predictions are different from those expected by the classical flux-flow model in which linear regimes in I-V characteristics at $I > I_C$ are magnetic field dependent ($dV/dI \sim H$) [13,14].

Finally, going back to the Eq. (2), as $I_N$ increases there is a value of $I_N$ in the I-V curve in which $I_N$ reaches $I_S$. At this point, $I = 2I_N = 2I_S$ or $I_S = I/2$ which provides (see Eq. 4)

$$V = ( R_N / 2 ) I \quad . \qquad (7)$$

In order to verify if the approaching of $I_N$ to $I_S$ is related to the linear regime of an I-V curve, let us compare Eqs. (5) and (7). This procedure yields $I = 2 I_C$ and, consequently, $I_N = I_S = I_C$. Such a result indicates that for values of $I = 2 I_C$ a crossing point separating the non-linear from the linear regimes takes place in the I-V characteristics for each applied magnetic field. In the regime where $I < 2I_C$, the behavior of I-V curves is dominated by the supercurrent component (non-linear regime). On the other hand, for $I > 2I_C$, the dissipation is dominated by the single particle current and the I-V characteristics display linear regimes for applied magnetic fields.

In order to verify the applicability of the RSJ model to a system comprised of many Josephson junctions, we describe here a systematic investigation of transport properties in polycrystalline samples exhibiting the double resistive superconducting transition. The investigated samples belong to the hole-doped $Y_{1-x}Pr_xBa_2Cu_3O_{7-\delta}$ (x = 0.0; 0.35, and 0.45) and electron-doped $Sm_{2-x}Ce_xCuO_{4-\delta}$ (x = 0.18) systems.

## II. EXPERIMENTAL PROCEDURE

Polycrystalline samples of the system $Y_{1-x}Pr_xBa_2Cu_3O_{7-\delta}$ were prepared by the solid state reaction method. The powders were mixed in appropriate amounts and sintered at different temperatures depending on sample composition. Phases were identified by X-ray powder diffractometry using Cu-K$\alpha$ radiation. Optical and scanning electron microscopy revealed the granular structure of the samples. These results showed that the samples are single phase. Details about these procedures were already reported elsewhere [12].

Polycrystalline samples of $Sm_{1.82}Ce_{0.18}CuO_{4-\delta}$ were prepared by using the sol-gel route [22]. Powder X-ray diffraction revealed single phase material with the T'-structure. Details about the sample preparation and the chemical reduction necessary to induce superconducting properties in this compound are given in Ref. 6 and 22.

All the samples were characterized by magnetotransport measurements. Copper electrical leads were attached to Au-film contact pads (contact resistance < 1 Ω) on parallelepiped-shaped samples using Ag epoxy. The applied electrical current was alternated in all measurements in order to eliminate the thermopower effects produced in the voltage contacts [23]. Conventional four-wire electrical resistance and magnetoresistance were performed in the limit of low applied magnetic fields, employing excitation current from 1 to 100 mA. The voltage versus excitation current I-V curves were performed for several values of applied magnetic field in the range $0 \leq H \leq 5000$ Oe, employing excitation current up to about 50 mA. Before each measurement, the samples were cooled down from the normal state in zero applied magnetic field after which the magnetic field was turned on.

### III. RESULTS AND DISCUSSION

Figure 2 shows the temperature dependence of the electrical resistance R(T) for $Y_{0.55}Pr_{0.45}Ba_2Cu_3O_{7-\delta}$ and $Sm_{1.82}Ce_{0.18}CuO_{4-\delta}$ samples measured at zero and under applied magnetic field for several values of excitation current. From the figure we see that the double superconducting transitions occur at temperatures $T_{ci}$ and $T_{cj}$. The regime between $T_{ci}$ and $T_{cj}$ is related to an ohmic behavior which has been attributed to an increasing of the volume fraction of superconducting clusters with decreasing temperature [24]. This is much more pronounced in $Y_{0.55}Pr_{0.45}Ba_2Cu_3O_{7-\delta}$ where a systematic decrease in R(T) between $T_{ci}$ and $T_{cj}$ occurs.

Since R(T) below $T_{cj}$ for both samples is excitation current dependent one is able to assure that a non-ohmic regime appears due to the increasing I. This is the expected behavior in a system comprised of superconducting clusters linked by Josephson coupling. In such a system increasing excitation current results in a suppression of the coupling between superconducting clusters leading the system to a non-zero resistance state at low temperatures.

In order to further investigate the dissipation at temperatures below $T_{cj}$, we have measured I-V characteristics for representative samples belonging to the $Y_{1-x}Pr_xBa_2Cu_3O_{7-\delta}$ and $Sm_{2-x}Ce_xCuO_{4-y}$ systems. The results of a set of I-V characteristics for these samples measured at fixed temperatures, for different values of applied magnetic fields are displayed in Fig. 3. From these curves it is clear that the I-V characteristics exhibit linear behavior at high applied excitation currents, $I > I_c$, which are independent of the material. $I_C$ was defined as the current value in which V = 0, obtained from the extrapolation of the linear regime of the V(I) curve down to lowest value of the excitation current. Most importantly the data displayed in Fig. 3 show that, in the high current limit, the slopes of the I-V characteristics are magnetic field independent. Such a feature is supported by the results shown in Fig. 4.

We concentrate now in the linear regime of the I-V characteristics at high applied excitation currents ($I \gg I_C$) for the $Y_{0.55}Pr_{0.45}Ba_2Cu_3O_{7-\delta}$ sample. Fig. 4(a) shows the first derivative of the I-V curves displayed in Fig. 3(a), at several values of applied magnetic fields in the range $0 \leq H \leq 76$ Oe. At high excitation currents, the dV/dI curves approach a constant value, a feature which is magnetic field independent. We better emphasize that the linear behavior of R(T) and its saturation under applied magnetic fields can not be explained within the context of the classical flux flow model. According to this model, the magnitude of dV/dI is strongly magnetic field dependent as described in details in Ref. 14.

In order to gain further information regarding the behavior of the transport properties of this system let us compare the saturation value of dV/dI at several applied magnetic fields (see Fig. 4(a)) with the magneto-resistance [$R(H) \equiv V(H) / I$] curves displayed in Fig. 4(b). These curves were obtained from measurements of R(H) when the system is subjected to different excitation currents at T = 4.2 K. The combined results indicates that Ohmic behavior is only observed for high applied magnetic fields (H > 50 Oe) and high excitation currents ($I \geq 40$ mA). In addition to this, the data also show that the electrical resistance of the sample reaches a saturation value close to 0.36 $\Omega$ in both situations. It is important to notice that this ohmic resistance is related to the normal state shunt resistance $R_N$, as displayed in Fig. 1. Such a normal state is characterized by disconnected superconducting clusters, in which the normal state shunt resistance of the network of junctions was reached. This idea is supported by previous magnetotransport data of granular superconductors in which dissipation was argued to be ruled by the weak intergranular coupling and by intragranular flux flow at low and high magnetic fields, respectively [24].

The values of $R_N$ are of interest for an analysis of the transport measurements discussed above and for a better understanding of the RSJ model. Within this context, Fig. 5 displays in detail the I-V characteristics taken at 4.2 K and for several values of applied magnetic fields in the range $0 \leq H \leq 76$ Oe. A number of important features can be identified in these curves. The first one is that with increasing H results in an appreciable reduction of the $I_C$ values as well as a progressive increase of the linear portion of the I-V curves. To gain further insight into the RSJ model, together with the I-V characteristics displayed at Fig. 5, we have plotted the linear curve predicted by Eq. (7), $V = (R_N/2)I$. For all H values, there is a crossing point that separates the non-linear from linear regime of the I-V curve. Such a point lies in the intersection of the $V = (R_N/2)I$ and I-V curves. Furthermore, such a crossing point occurs at the current value $I \sim 2I_C$, as expected from the RSJ model. For $H = 0$, for instance, $I_C \sim 33$ mA and the crossing point occurs at $I \sim 2I_C \sim 66$ mA. In high excitation current limit ($I > 2I_C$), all I-V curves are linear and exhibit a sloop very close to $R_N \sim 0.36$ Ω. It must be stressed that the results described above were observed in all I-V characteristics (not shown) taken at several temperatures $T < T_{cj}$. These results indicate that the transport properties of the investigated samples are in complete agreement with the predictions of the RSJ model.

Going further in the RSJ model, we have estimated the excitation current dependence of both the supercurrent ($I_S$) and the normal current ($I_N$) for the $Y_{0.55}Pr_{0.45}Ba_2Cu_3O_{7-\delta}$ compound. In order to do this we have used the $R_N$ value (0.36 $\Omega$) obtained for this compound (see Fig. 4) as well as the Eq. (2) and (3). Representative $I_S(I)$ and $I_N(I)$ curves estimated from the I-V characteristics for $Y_{0.55}Pr_{0.45}Ba_2Cu_3O_{7-\delta}$ at 4.2 K and under H = 2.85 Oe (see Fig. 5(a)) are displayed in Fig. 6. From these curves one is able to see that the excitation current coincides with the supercurrent for low values of excitation current. With increasing applied current, $I_S$ reaches $I_C$ at $I = 2I_C$ ($I_S = I_N = I_C$). Further increase in the excitation current leads $I_S$ to exhibit a saturation value. At this point, the excitation current dependence of the voltage is linear, and $I = I_N + I_C$, where $I_C$ is the maximum supercurrent which crosses the network of junctions.

Based on the RSJ model and the experimental results above described we argue that the dissipation due to the normal current occurs in all applied current range studied. In fact, it vanishes at very low excitation current due to the detection limit of the voltage instruments. Thus, when I $\rightarrow$ zero and I $\ll$ $I_C$, $I_S \gg I_N$, and I $\sim$ $I_S$, which describes a zero-resistance (V $\rightarrow$ 0) superconducting state. For instance, at T = 4.2 K, H = 2.85 Oe, and I = $10^{-4}$ mA, $I_S \sim 10^{-4}$ mA and $I_N \sim 10^{-6}$ mA, indicating that $I_N$ is nearly hundred times lower than $I_S$ (see inset of the Fig. 6).

Finally, we have analyzed all the I-V characteristic curves for $Sm_{1.82}Ce_{0.18}CuO_{4-\delta}$, $YBa_2Cu_3O_{7-\delta}$, and $Y_{0.65}Pr_{0.35}Ba_2Cu_3O_{7-\delta}$ compounds using a similar procedure as described in detail for the $Y_{0.55}Pr_{0.45}Ba_2Cu_3O_{7-\delta}$. Similar results were obtained from the application of the RSJ model to these samples

## IV. CONCLUSION

In summary, we have investigated the transport properties of polycrystalline samples of both the hole-doped $Y_{1-x}Pr_xBa_2Cu_3O_{7-\delta}$ (x = 0.0, 0.35, and 0.45) and the electron-doped $Sm_{2-x}Ce_xCuO_{4-\delta}$ (x = 0.18) systems. All the investigated samples exhibit the double resistive superconducting transition, a feature characterized by drops in the electrical resistivity data at two well defined temperatures $T_{ci}$ and $T_{cj}$. We have found that, in the limit of high excitation current, the I-V characteristics taken at T below $T_{cj}$ are linear for all the investigated samples. Most importantly it was also found that the slopes of the I-V characteristics at this high-current limit are magnetic field independent. The results obtained for four different samples are in good agreement with the predictions of the RSJ model suggesting that dissipation in granular samples is a consequence of the normal current flowing in parallel with the supercurrent.

## ACKNOWLEDGEMENTS

This work has been supported by the Brazilian agencies FAPESP under Grant No. 00/03610-4 and 97/11113-6, CNPq, and CAPES. The authors are grateful to B. Ferreira for technical assistance. Work at Universidade de São Paulo was supported by FAPESP under Grant No. 05/53241-9. R. F. Jardim is a CNPq fellow under Grant No. 303272/2004-0.

**FIGURE CAPTIONS**

**Figure 1 -** Schematic diagram of the resistively shunted Josephson junction model. $R_N$ and JJ are the shunt resistance and the Josephson junction, respectively.

**Figure 2** – Temperature dependence of the electrical resistance for the (a) $Y_{0.55}Pr_{0.45}Ba_2Cu_3O_{7-\delta}$ and (b) $Sm_{1.82}Ce_{0.18}CuO_{4-\delta}$ samples measured in different applied current and zero applied magnetic field. The insets show similar curves but under an applied magnetic field.

**Figure 3 -** I-V characteristics for the (a) $Y_{0.55}Pr_{0.45}Ba_2Cu_3O_{7-\delta}$ and (b) $Sm_{1.82}Ce_{0.18}CuO_{4-\delta}$ samples measured at several values of applied magnetic field and at T = 4.2 K. In the insets are shown similar results obtained at high excitation currents: $YBa_2Cu_3O_{7-\delta}$ (top) and $Y_{0.65}Pr_{0.35}Ba_2Cu_3O_{7-\delta}$ (bottom) at 77 and 34.5 K, respectively.

**Figure 4** – (a) First derivative of the V versus I curves for the $Y_{0.55}Pr_{0.45}Ba_2Cu_3O_{7-\delta}$ sample (displayed in Fig. 3(a)). (b) Magnetoresistance curves taken at T = 4.2 K when the sample was subjected to several values of the excitation current.

**Figure 5** – Selected I-V characteristics under applied magnetic fields for the $Y_{0.55}Pr_{0.45}Ba_2Cu_3O_{7-\delta}$ sample. $R_N$ and $I_C$ are defined in the text.

**Figure 6 -** Excitation current I dependence of both the supercurrent ($I_S$) and the normal current ($I_N$). The solid lines were calculated by using Eqs. (2) and (3) and data of I-V characteristics for the compound $Y_{0.55}Pr_{0.45}Ba_2Cu_3O_{7-\delta}$ taken at T = 4.2 K and under H = 2.85 Oe. The inset shows the same curves in a log-log scale.

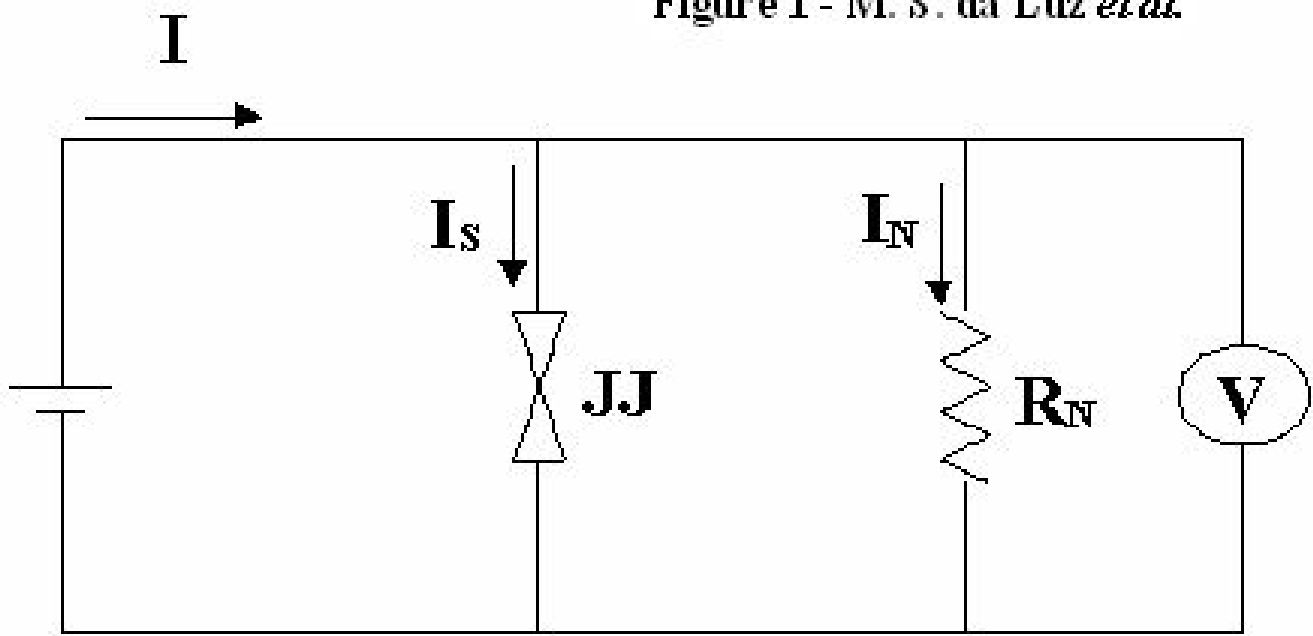

Figure 1 - M. S. da Luz *et al.*

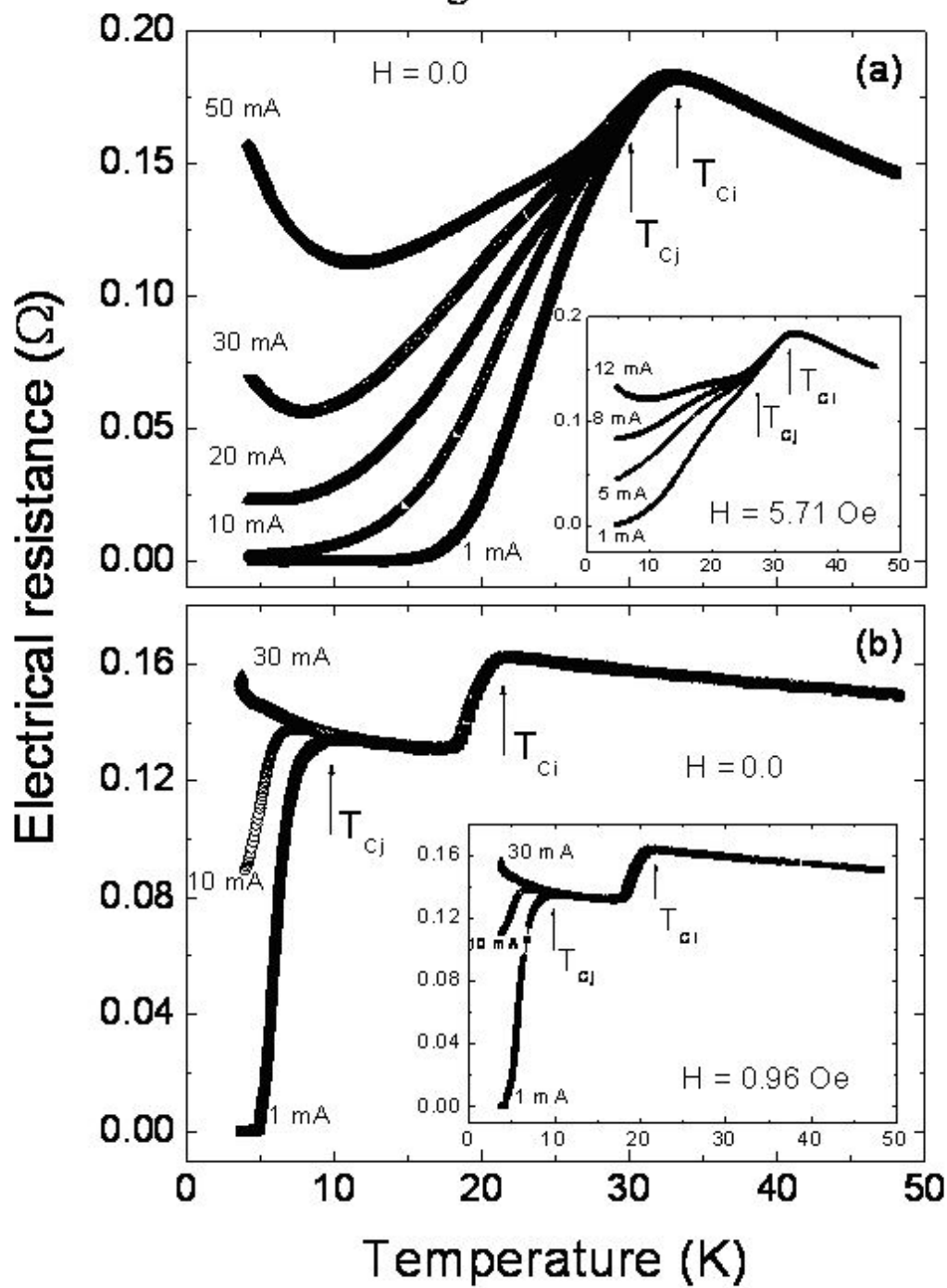

Figure 2 - M. S. da Luz et al.

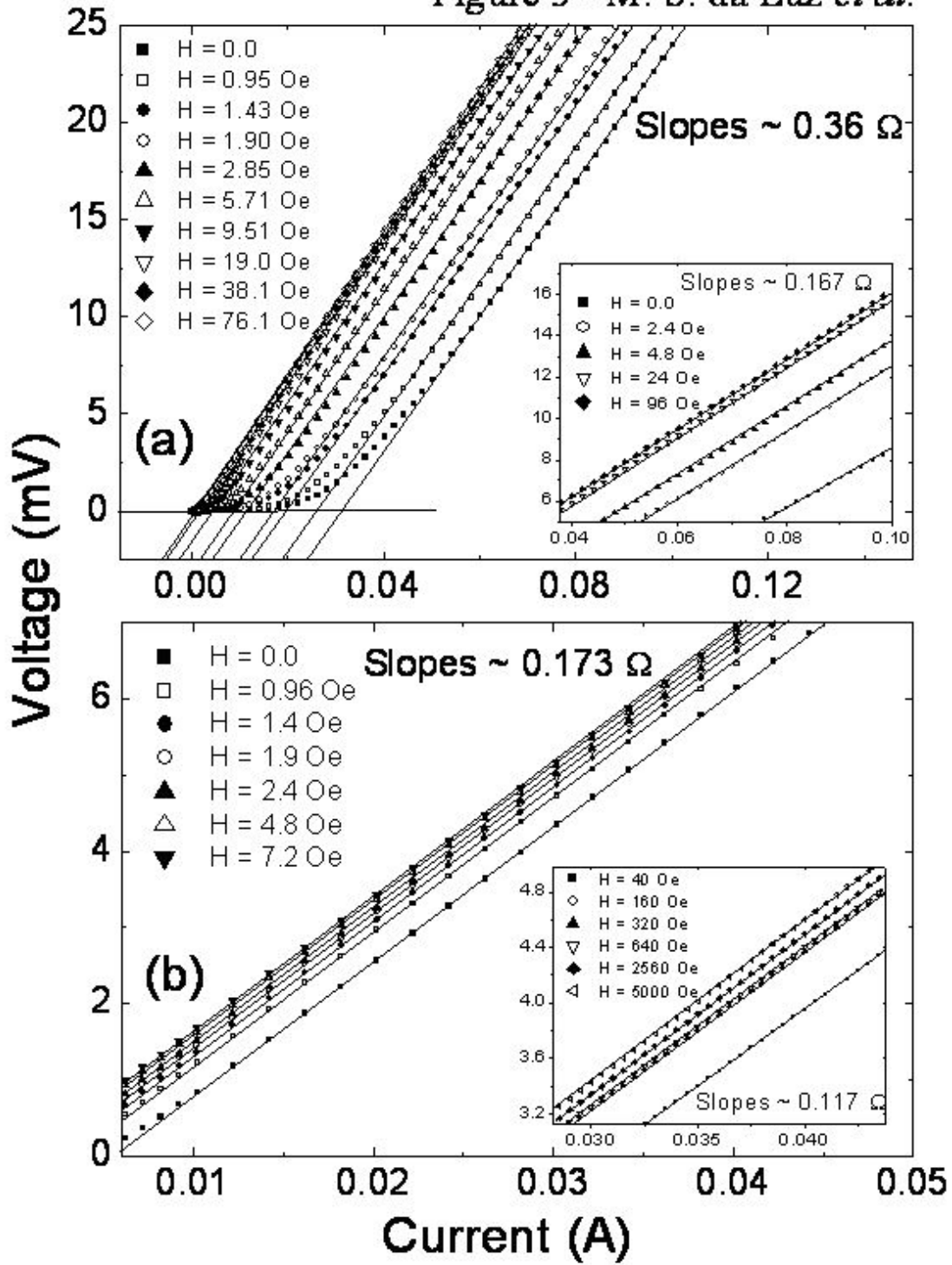

Figure 3 - M. S. da Luz et al.

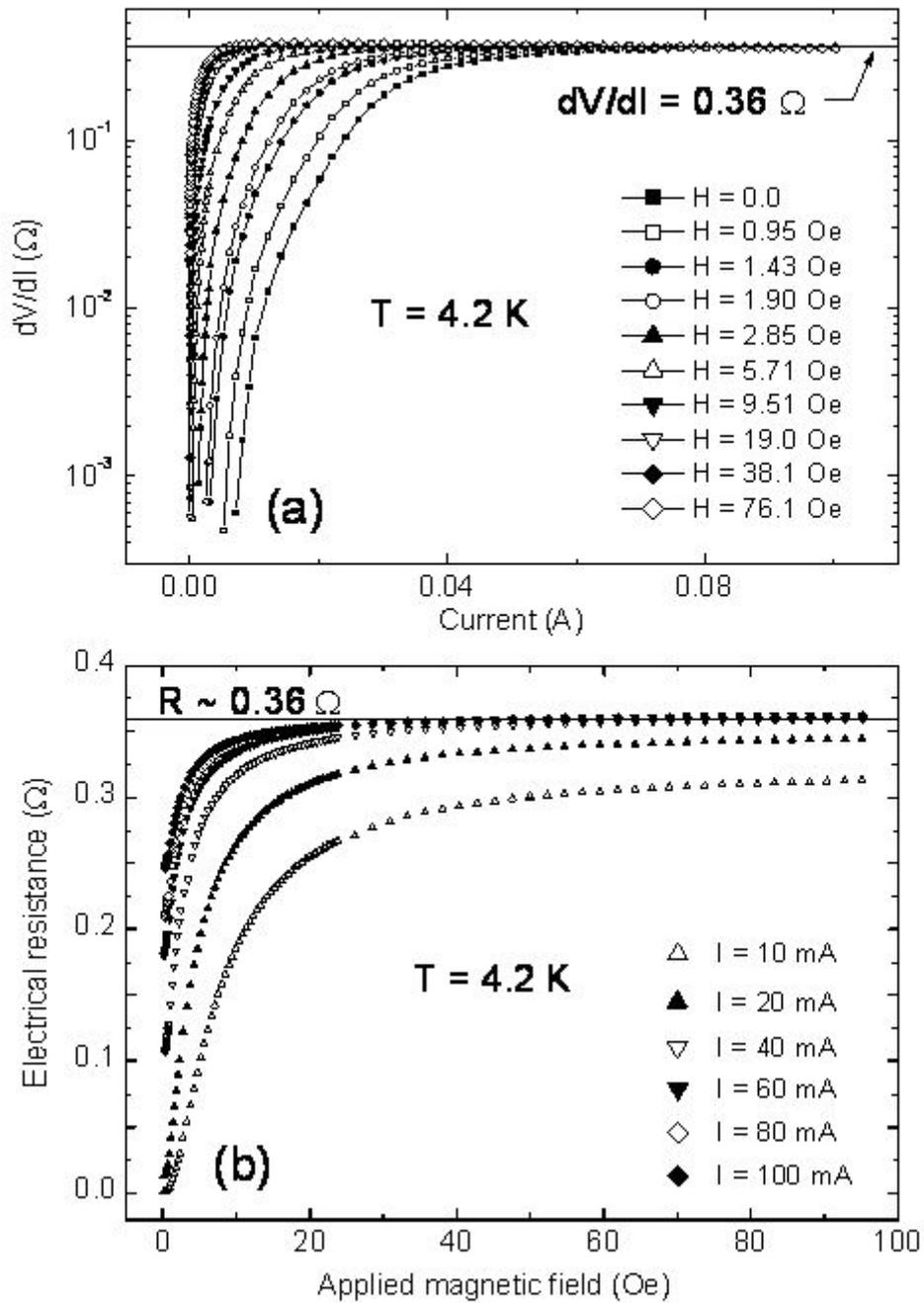

Figure 4 - M. S. da Luz *et al*.

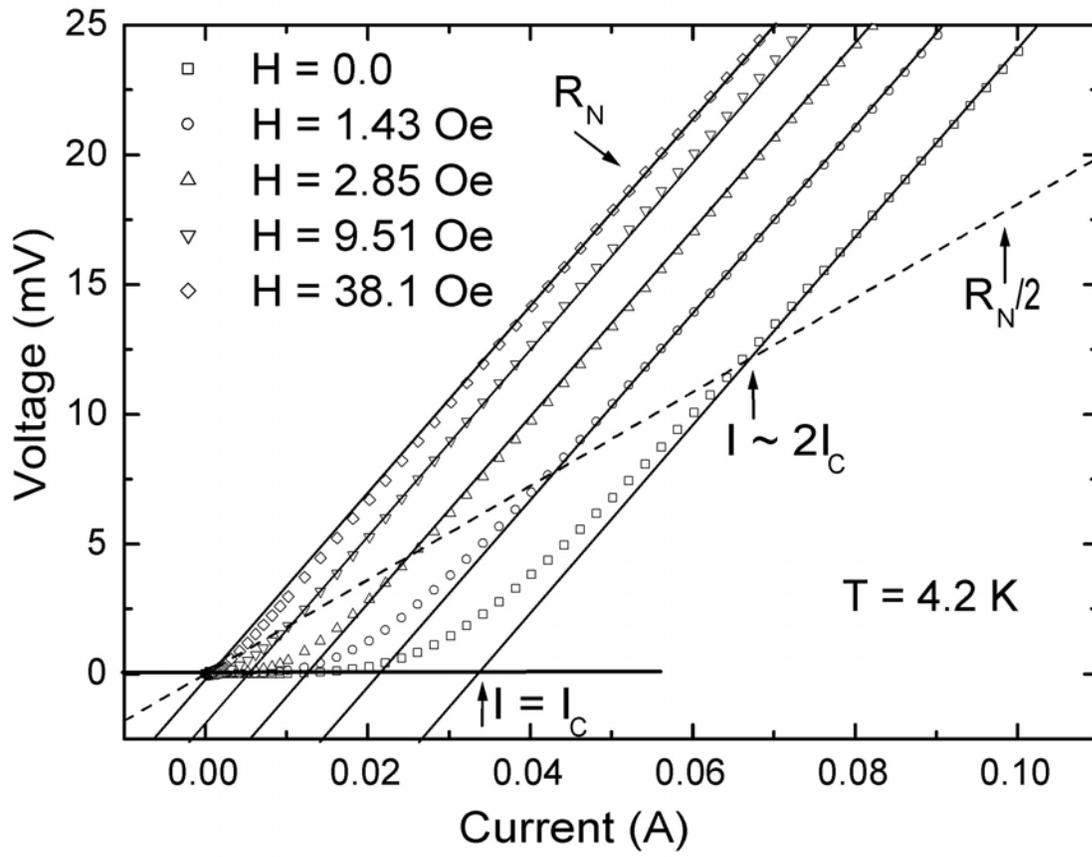

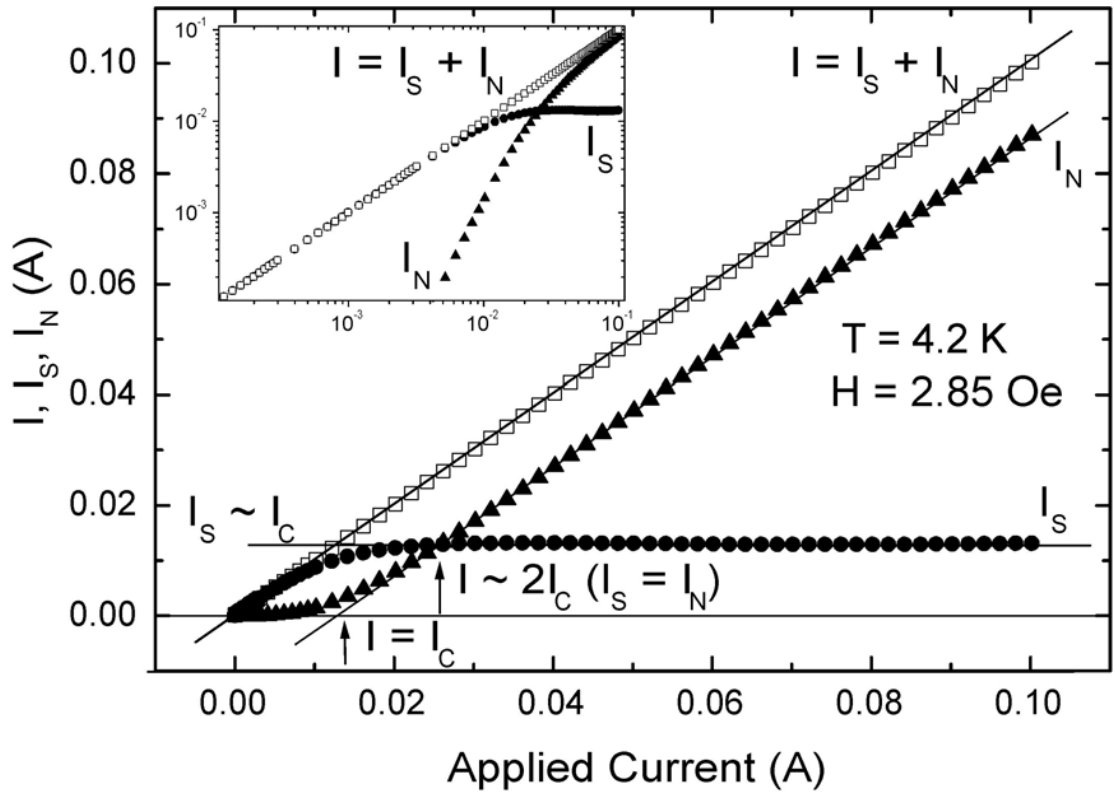